\documentclass{ws-mpla}
\usepackage{graphicx,cite}
\usepackage[usenames,dvipsnames]{color} 
\usepackage{slashed}

\newcommand{\newc}{\newcommand}
\newc{\renewc}{\renewcommand}

\def\beq{\begin{equation}}
\def\eeq{\end{equation}}
\def\bea{\begin{eqnarray}}
\def\eea{\end{eqnarray}}
\def\bitem{\begin{itemize}}
\def\eitem{\end{itemize}}
\def\ba{\begin{array}}
\def\ea{\end{array}}
\def\bal{\begin{align}}
\def\eal{\end{align}}
\def\bi{\begin{itemize}}
\def\ei{\end{itemize}}
\def\lsim{\mathrel{\rlap{\lower4pt\hbox{\hskip1pt$\sim$}}
    \raise1pt\hbox{$<$}}}         
\def\gsim{\mathrel{\rlap{\lower4pt\hbox{\hskip1pt$\sim$}}
    \raise1pt\hbox{$>$}}}
\newc{\blue}{\textcolor{blue}}
\newc{\red}{\textcolor{red}}

\newc{\ie}{{\it i.e.~}}          \newc{\etal}{{\it et al.~}}
\newc{\eg}{{\it e.g.~}}          \newc{\etc}{{\it etc.~}}
\newc{\cf}{{\it c.f.~}}
\newc{\os}{\mbox{\hspace{4pt}}}
\newc{\us}{\mbox{\hspace{12pt}}}

\renewc{\bar}{\overline}

\newc{\gev}{\,{\rm GeV}}
\newc{\mev}{\,{\rm MeV}}
\newc{\ev}{\,{\rm eV}}
\newc{\kev}{\,{\rm keV}}
\newc{\tev}{\,{\rm TeV}}

\newc{\LM}{\mathcal{L}}
\newc{\SM}{\mathcal{S}}

\newc{\HM}{\mathcal{H}}
\newc{\GM}{\mathcal{G}}
\newc{\OM}{\mathcal{O}}
\newc{\FM}{\mathcal{F}}
\newc{\AM}{\mathcal{A}}
\newc{\BM}{\mathcal{B}}
\newc{\NM}{\mathcal{N}}
\newc{\WM}{\mathcal{W}}
\newc{\ZM}{\mathcal{Z}}

\newc{\Chi}{\mathcal{X}}
\newcommand{\met}{\slashed{E}_T}

\begin{document}

\markboth{T. Flacke, K. C. Kong, S. C. Park}
{Non minimal UED: a review}

\catchline{}{}{}{}{}

\title{A Review on Non-Minimal Universal Extra Dimensions}

\author{Thomas Flacke}

\address{Department of Physics, \\Korea Advanced Institute of Science and Technology, \\
335 Gwahak-ro, Yuseong-gu, Daejeon 305-701, Korea\\
flacke@kaist.ac.kr}

\author{Kyoungchul Kong}

\address{Department of Physics and Astronomy, University of Kansas, \\Lawrence, KS 66045 USA
\\
kckong@ku.edu}

\author{Seong Chan Park}

\address{Department of Physics, Sungkyunkwan University, \\Suwon 440-746, Korea\\
and\\
Korea Institute for Advanced Study, \\Seoul 130-722, Korea \\
s.park@skku.edu}

\maketitle


\begin{abstract}
We report on the current status of non-minimal universal extra dimension (UED) models. 
Our emphasis is on the possible extension of the minimal UED model by allowing bulk masses and boundary localized terms. 
We  take into account the data from the Large Hadron Collider (LHC) as well as direct and indirect searches of dark matter and electroweak precision measurements.
\end{abstract}
 
\section{Introduction}

\begin{figure}[t]
\centerline{\includegraphics[width=.9\textwidth]{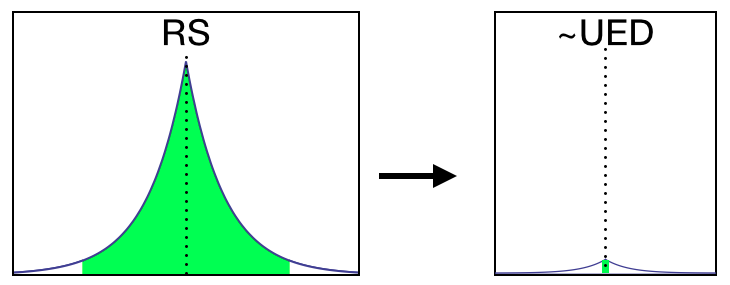}}
\vspace*{8pt}
\label{fig:RSUED}
\caption{An effective description of a RS model after integrating out the part in the vicinity of UV could be regarded as a UED model.}
\end{figure}

Universal Extra Dimension (UED) refers to models with (nearly) flat extra dimensions in which all the standard model fields can propagate \cite{Appelquist:2000nn}. A field in higher dimensions is expanded into its Fourier modes, so called Kaluza-Klein (KK) modes.
Its lowest mode or the zero mode is regarded as one of the standard model (SM) particles, i.e. quarks, leptons, gauge bosons and the Higgs boson which have been observed at the LHC. If the extra dimension is `symmetric' under the reflection about its midpoint, which may be regarded as the consequence of two throats compactification~\cite{Csaki:2010az} as shown in Fig. \ref{fig:RSUED}, a KK mode carries a definite parity (even or odd), dubbed KK-parity.
It is assigned as $P_n=(-1)^n$ for a $n$-th KK mode. 
Since a standard model particle is a zero mode and has an even parity, the lightest KK-parity odd particle (LKP) becomes automatically stable as it can never decay into only zero modes due to parity conservation. Thanks to this property (stability of the LKP), UED models are often regarded as a theory of dark matter (DM). The first excited state of the $U(1)_Y$ gauge boson, $B_1$,  is the best studied candidate of KK dark matter in the so-called minimal UED model (MUED). 
See Refs. \cite{Hooper:2007qk,Kong:2010mh,Datta:2010us,Servant:2014lqa} for recent reviews on MUED. 

In this review, we focus on the recently proposed extensions of the minimal model and their status after LHC8 and several dark matter detection experiments. 
Two notable extension in 5D (and their variations) have been discussed in literature:
\begin{itemize}
\item {Split-UED\cite{Park:2009cs,Chen:2009gz,Chen:2009zs,Kong:2010xk,Kong:2010qd}, $M_5\neq 0$}
\item {non-minimal UED\cite{Flacke:2008ne,Datta:2012tv,Flacke:2013pla,Flacke:2013nta,Datta:2013yaa}, $r \neq 0$  (and $M_5\neq 0)$} 
\end{itemize}
where non-vanishing bulk masses for fermions ($\sim M_5(y) \bar{\Psi}\Psi$) and boundary localized operators ($\sim r \left[\delta(y- 0)+\delta(y-L)\right] {\cal O}$) at boundaries $y=0, L$ are introduced such that KK-parity is respected and therefore KK dark matter is still available. In particular the inclusion of bulk masses and boundary localized terms is natural as such terms are allowed by all symmetries of the model.  Even in their absence of boundary localized terms at tree-level, they are induced at loop-level \cite{Cheng:2002iz}. Therefore, the assumption of their presence does not require additional structure in the theory. 
To the opposite, It is rather unnatural if one assumes the absence of all boundary localized operators at one scale. 
For further extensions of UED model including embedding in higher dimensions, 
we refer to Refs.\cite{Dobrescu:2001ae,Appelquist:2001mj,Dobrescu:2004zi,Burdman:2005sr,Ponton:2005kx,Burdman:2006gy,Dobrescu:2007ec,Dobrescu:2007xf,Nishiwaki:2011vi,Choudhury:2011jk,Dohi:2010vc,Cacciapaglia:2009pa,Arbey:2012ke}.

\section{Non-minimal UED}\label{sec:nmUED}

In this section, we briefly review Non-minimal (or Next-to-Minimal) UED (NMUED) with non-vanishing bulk masses and boundary localized terms. 
We follow notations and conventions in Ref. \cite{Flacke:2013pla}. The action is invariant under the standard model gauge symmetry 
$ G_{\rm SM}=SU(3)_c \times SU(2)_W \times U(1)_Y$ as well as KK-parity and  consists of two parts:
\begin{enumerate}
\item The bulk action $S_5$ (invariant under 5 dimensional Lorentz symmetry)
\item The boundary action $S_{bdy} \left[\delta(y)+\delta(y-L)\right]$ (invariant under 4 dimensional Lorentz symmetry)
\end{enumerate}
The 5 dimensional (5D) fermion fields (with a possible extension with the right handed neutrino for the non-vanishing neutrino mass) are introduced with their gauge charges as follows, 
\begin{eqnarray}
&&Q = (3,2)_{1/6}\ni Q_L^{(0)} = \binom{U_L^{(0)}}{D_L^{(0)}}, 
U =(3,1)_{2/3} \ni U_R^{(0)}, 
D =(3,1)_{-1/3} \ni D_R^{(0)}, \\
&&L= (1,2)_{-1/2}\ni L_L^{(0)} = \binom{\nu_L^{(0)}}{e_L^{(0)}}, 
E=(1,1)_{-1} \ni e_R^{(0)} , ~~(N =(1,1)_{0} \ni \nu_R^{(0)} ) \, , \nonumber 
\end{eqnarray}
where the superscript $^{(0)}$ denotes the zero mode of the KK tower.
The bulk action ($S_5$) is given by 
\beq
 S_5=\int d^4x \int_{-L}^L dy  \, \left[ {\cal L}_V+{\cal L}_{\Psi}+{\cal L}_H+{\cal L}_{Yuk}\right] ,
\label{5Daction}
\eeq
where  
\begin{eqnarray}
&&{\cal L}_V= \sum_{\AM}^{G,W,B} -\frac{1}{4} \AM^{MN}\cdot \AM_{MN} \, , \\
&&{\cal L}_{\Psi}= \sum_{\Psi}^{Q,U,D,L,E}i \overline{\Psi} \overleftrightarrow{D}_M \Gamma^M \Psi - M_\Psi \overline{\Psi}\Psi \label{FLag}
\end{eqnarray}
where $\AM$ denotes the gauge bosons, gluon ($G$), weak gauge bosons ($W$) and the hypercharge gauge boson ($B$).
They appear in the gauge covariant derivatives, for instance $D_M = \partial_M +i g^5_s \lambda\cdot G_M
+  i g^5_w  \tau \cdot W_M  +  i g^5_Y Y B_M$ for quark doublets. The five dimensional coupling constants are denoted by $g^5_i$s and the symmetry generators are $\lambda$s and $\tau$s for $SU(3)_c$ and $SU(2)_W$, respectively. 
The fermion kinetic term is defined as 
$\overline{\Psi} \overleftrightarrow{D}_M \Psi =\frac{1}{2}\{ \overline{\Psi} (D_M \Psi) - (D_M \overline{\Psi}) \Psi \} $. 
The gamma matrices in five dimensions are $\Gamma^M = (\gamma^\mu, i \gamma_5)$, which satisfy $\{\Gamma^A,\Gamma^B\}=2\eta^{AB}=2{\rm diag}(1,-1,-1,-1,-1)$. 
The bulk mass term is chosen to be odd under the inversion: $M_\Psi(y) = -M_\Psi(-y)$. 
The five dimensional Lagrangian for the Higgs and Yukawa interactions is 
\bea
{\cal L}_H&=& \left(D_M H\right)^\dagger D^M H -V(H), \\
V(H)&=&- \mu_5^2 |H|^2+\lambda_5 |H|^4\,,\label{SHiggs}\\
{\cal L}_{Yuk}&=& \lambda_5^E\overline{L}HE + \lambda_5^D\overline{Q}HD+ \lambda_5^U\overline{Q}\tilde{H}D+\mbox{h.c.}\,,
\eea
where $\tilde{H}=i\tau_2H^*$.

The KK-parity conserving boundary terms, that are allowed by gauge invariance and 4 dimensional Lorentz symmetry are
\beq
S_{bdy}= \int d^4 x \int_{-L}^L d y  \, \left({\cal L}_{\partial V}+{\cal L}_{\partial \Psi}+{\cal L}_{\partial H}+{\cal L}_{\partial Yuk}\right) \left[\delta(y-L)+\delta(y+L)\right],
\eeq
where
\bea
{\cal L}_{\partial{V}} &=& \sum_{\AM}^{G,W,B} -\frac{r_\AM}{4}   \AM_{\mu\nu} \cdot \AM^{\mu\nu}, \\
{\cal L}_{\partial \Psi}&=& \sum_{\Psi=Q,L} i  r_{\Psi} \overline{\Psi}_L D_\mu \gamma^\mu \Psi_L  
+\sum_{\Psi=U,D,E} i  r_{\Psi}\overline{\Psi}_R D_\mu \gamma^\mu \Psi_R ,\\
{\cal L}_{\partial H}&=&r_H \left(D_\mu H\right)^\dagger D^\mu H+ r_\mu \mu_5^2 |H|^2-r_\lambda \lambda_5 |H|^4\,,\label{SHiggsbdy}   \label{eq:boundaryhiggs} \\
{\cal L}_{\partial Yuk}&=&r_{\lambda^E} \lambda_5^E\overline{L}HE+ r_{\lambda^D} \lambda_5^D\overline{Q}HD+ r_{\lambda^U} \lambda_5^U\overline{Q}\tilde{H}D+\mbox{h.c.}\,. \label{eq:boundaryyukawa}
\eea

Above general KK-parity preserving 5D UED model contains many new parameters in addition to SM parameters ($g_\mathcal{A}, \mu_5, \lambda_5, 
\lambda_5^{U,D,E}$), 
\begin{itemize}
\item $L=\pi R/2, \Lambda$ (MUED parameters)
\item $M_\Psi$ (bulk masses in Split UED)
\item $r_\AM, r_\Psi,r_H, r_\mu, r_\lambda, r_{\lambda^{U,D,E}}$ (boundary localized terms)
\end{itemize}
Simplifying assumptions are typically introduced in order to avoid the tree level FCNC problem and non-trivial mixings. These include a universal boundary parameter $r\equiv r_{Q,U,D,L,E}=r_{G,W,B}=r_{H,\mu,\lambda}=r_{\lambda^{U,D,E}}$ and a universal KK-odd fermion bulk mass $\mu \theta(y)=M_{Q,L}=-M_{U,D,E}$ where $\theta(y)$ is the step function. 
Generically, without any fine-tuning, $r \sim L$ and $\mu \sim L^{-1}$ ($\frac{r}{L} \sim \mu L  \sim {\cal O}(1)$) is expected, since they are allowed by all symmetries of the model. 
 In general, the cutoff scale ($\Lambda \gsim {\cal O}(10)/L$) is also a parameter but as shown in literature, the dependence on the cutoff in masses and couplings is usually logarithmic and leads to subdominant effects due to the low cutoff scale.

The standard procedure of performing the KK decomposition is as follows:
\begin{enumerate}
\item  Derive the 5D equation of motion from the quartic terms of the bulk action. Electroweak symmetry breaking will be taken into account in the later steps as a correction.
\item  Separate the 5 dimensional equation of motion into a 4 dimensional part ($x^\mu$-dependent part) and an extra dimensional part ($y$-dependent part).  The $y$-dependent part may be regarded as Fourier bases for KK expansion and the $x$ dependent parts as Fourier coefficients. A 5D fermion is expanded as
$$\Psi(x,y)=\sum_n \left(\psi_L^{(n)}(x)f_n^L(y) +\psi_R^{(n)}f_n^R(y)\right),$$
reflecting the fact that minimal spinor in 5D is vectorlike. 
Both vector and scalar bosons can be expanded by a common set of Fourier bases as
$$A_\mu(x,y)=\sum_n A_\mu^{(n)}(x)f_n^A(y),$$
and
$$\Phi(x,y)=\sum_n \phi^{(n)}(x)f_n^A(y) \, ,$$
because the bulk equation motion is identical for bosons.
\item  Solve the $y$-dependent wave functions and determine KK masses using boundary conditions at $y=\pm L$.
\item Determine the normalization factor by making the KK mode kinetic terms canonical.
\item Enter the KK decomposition into the 5D action and integrate over the extra dimension in order to determine interactions between the Kaluza-Klein modes 
$\psi^{(n)}_{L,R}$, $A^{(n)}_\mu$, $\phi^{(n)}$.\footnote{Contributions to masses from electroweak symmetry breaking generically require to re-diagonalize the KK gauge boson and fermion mass matrices before determining the interactions in the true mass eigenbasis.}
\end{enumerate}

\begin{figure}[t]
\centerline{\includegraphics[width=.97\textwidth]{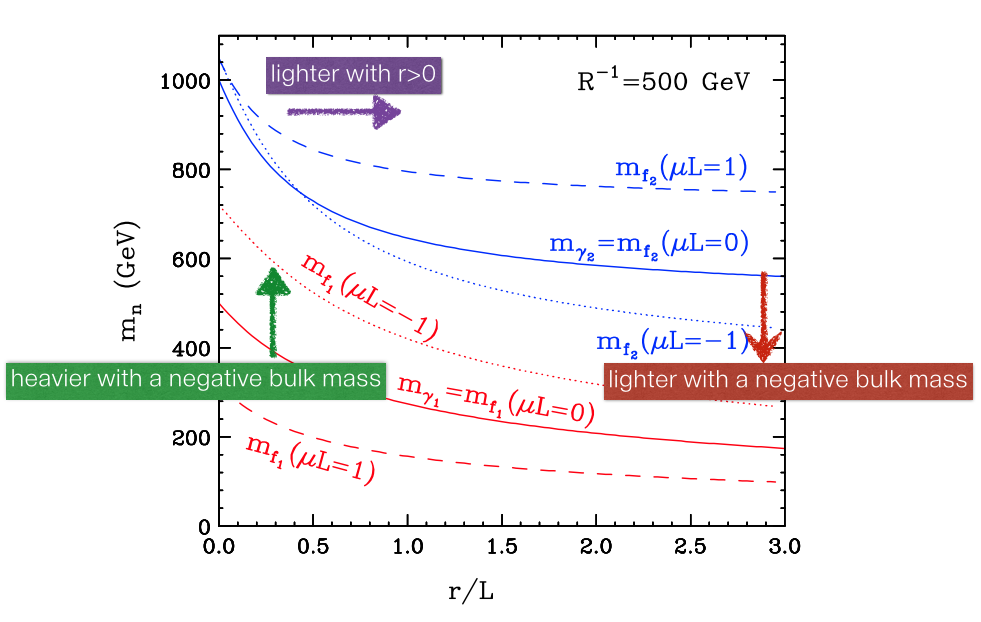}}
\vspace*{5pt}
\label{fig:kkmass}
\caption{KK masses at level-1 and level-2. Taken from Ref. \cite{Flacke:2013pla}.}
\end{figure}

The KK spectra and the coupling constants with the SM fields are all given in Ref.  \cite{Flacke:2013pla}. 
Here we summarize some noticeable features of KK-spectrum as shown in Fig. \ref{fig:kkmass} in the presence of universal parameters $r$ and $\mu$:
\begin{itemize}
\item For a non-zero boundary term, both KK fermions and KK bosons become lighter than those for $r=0$ ($m_{KK}(r=0)<m_{KK}(r>0)$). 
\item A larger $r$ means a lighter KK particle, $m_{KK} (r) < m_{KK} (r^\prime)$ if $r > r^\prime$.
\item If the bulk mass of a fermion is zero ($\mu L=0$), masses of KK fermion and KK boson are identical, $m_{f_{KK}}=m_{A_{KK}}$ for any $r$ (up to electroweak symmetry breaking effects).
\item 
A negative bulk mass  ($\mu L <0$) raises the masses of odd-numbered fermion KK modes while it lowers the masses of even-numbered fermion KK modes. The opposite holds for a positive bulk mass.
\end{itemize}

\begin{figure}[t]
\centering
\centerline{ 
\epsfig{file=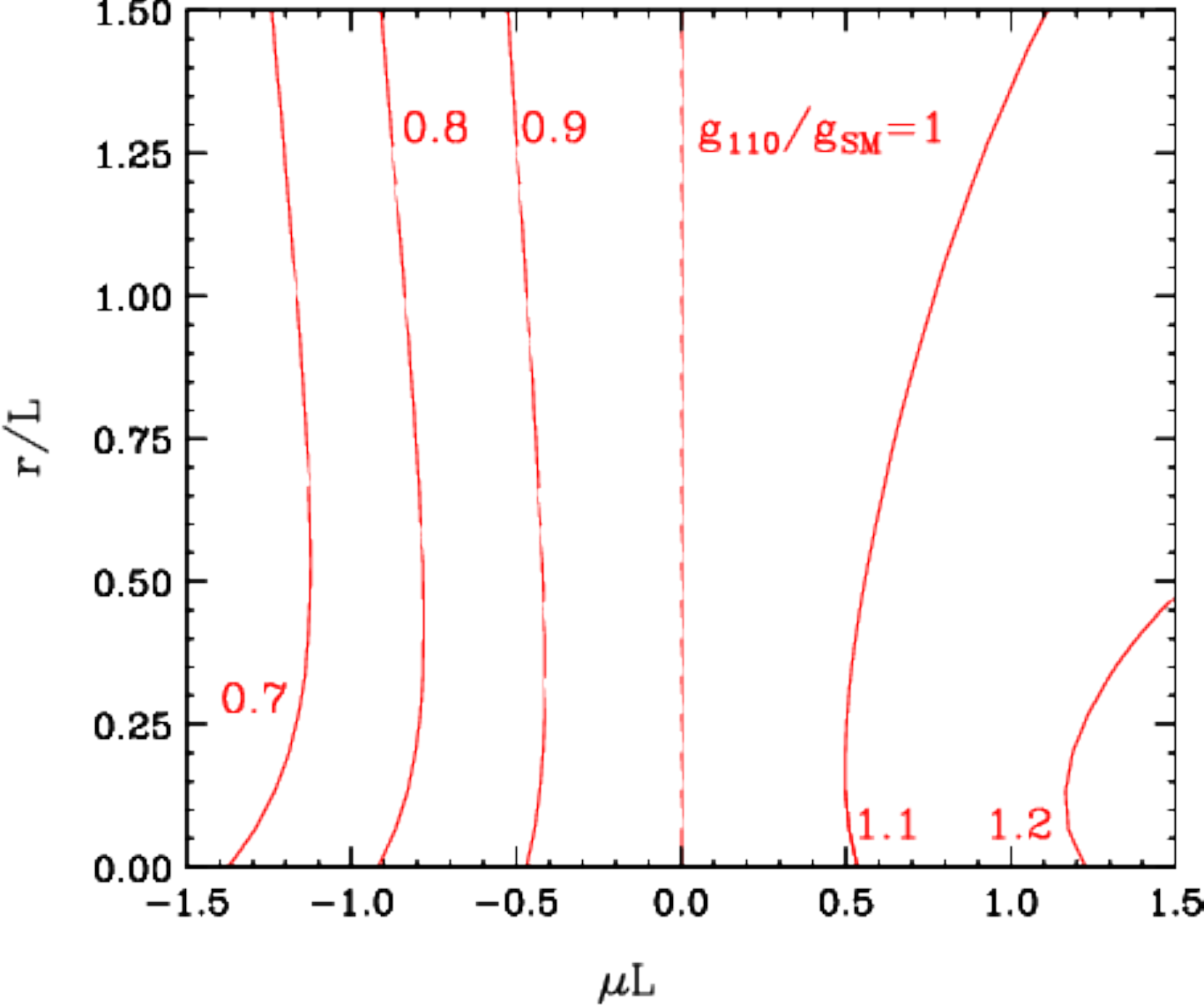, width=0.48\columnwidth} 
\hspace{0.1cm}
\epsfig{file=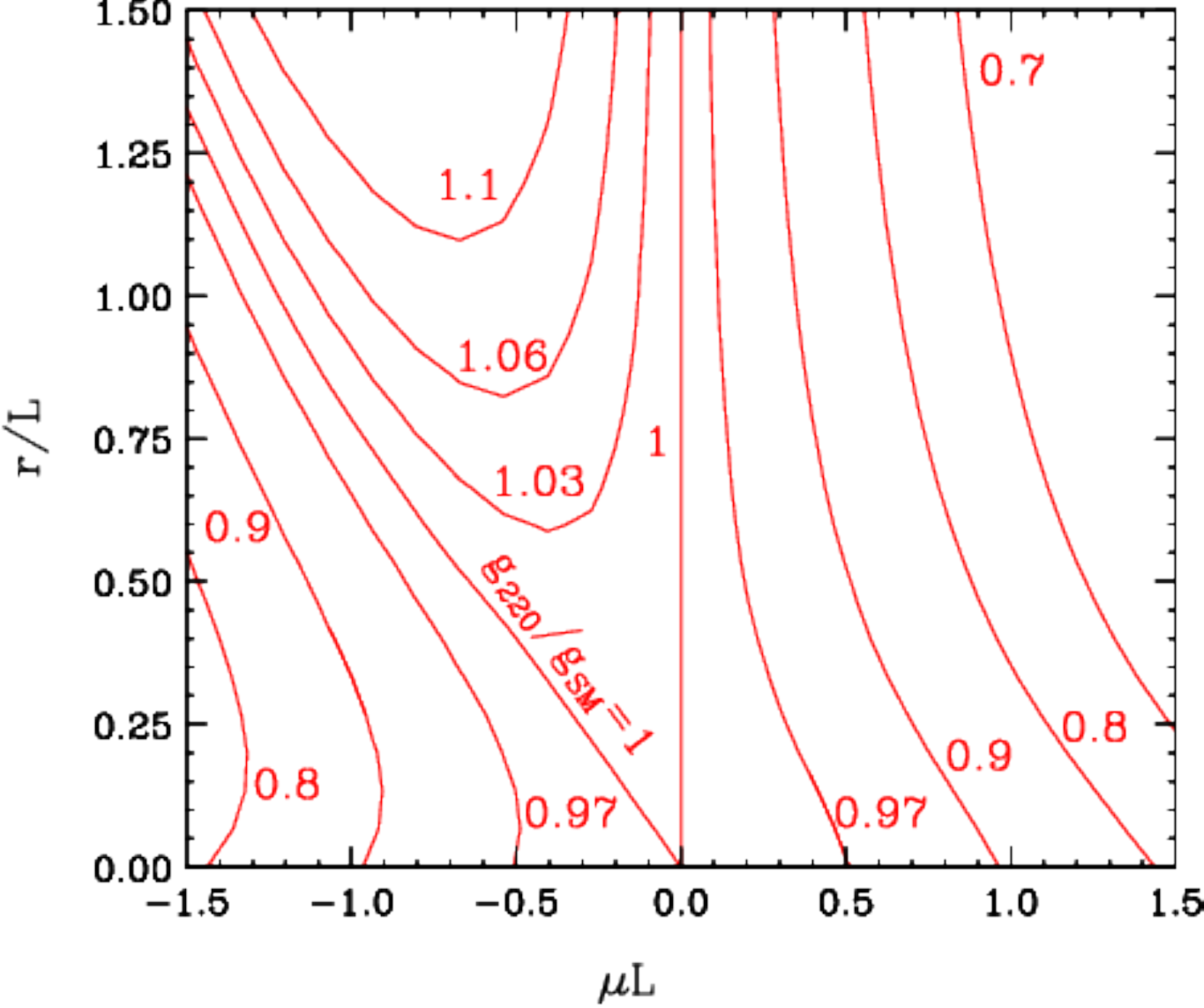, width=0.48\columnwidth}
\vspace{0.3cm}
}
\centerline{ 
\epsfig{file=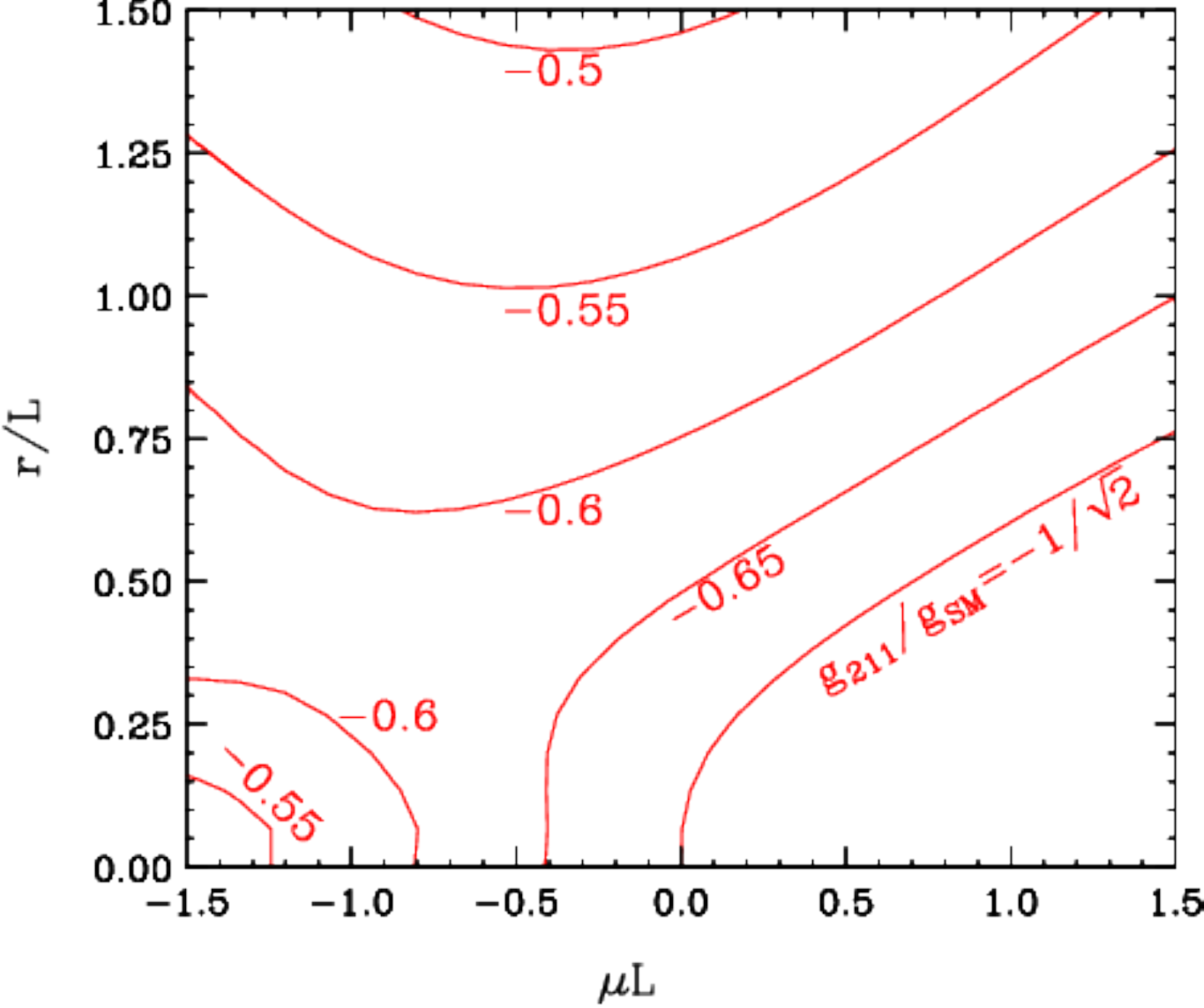, width=0.48\columnwidth} 
\hspace{0.1cm}
\epsfig{file=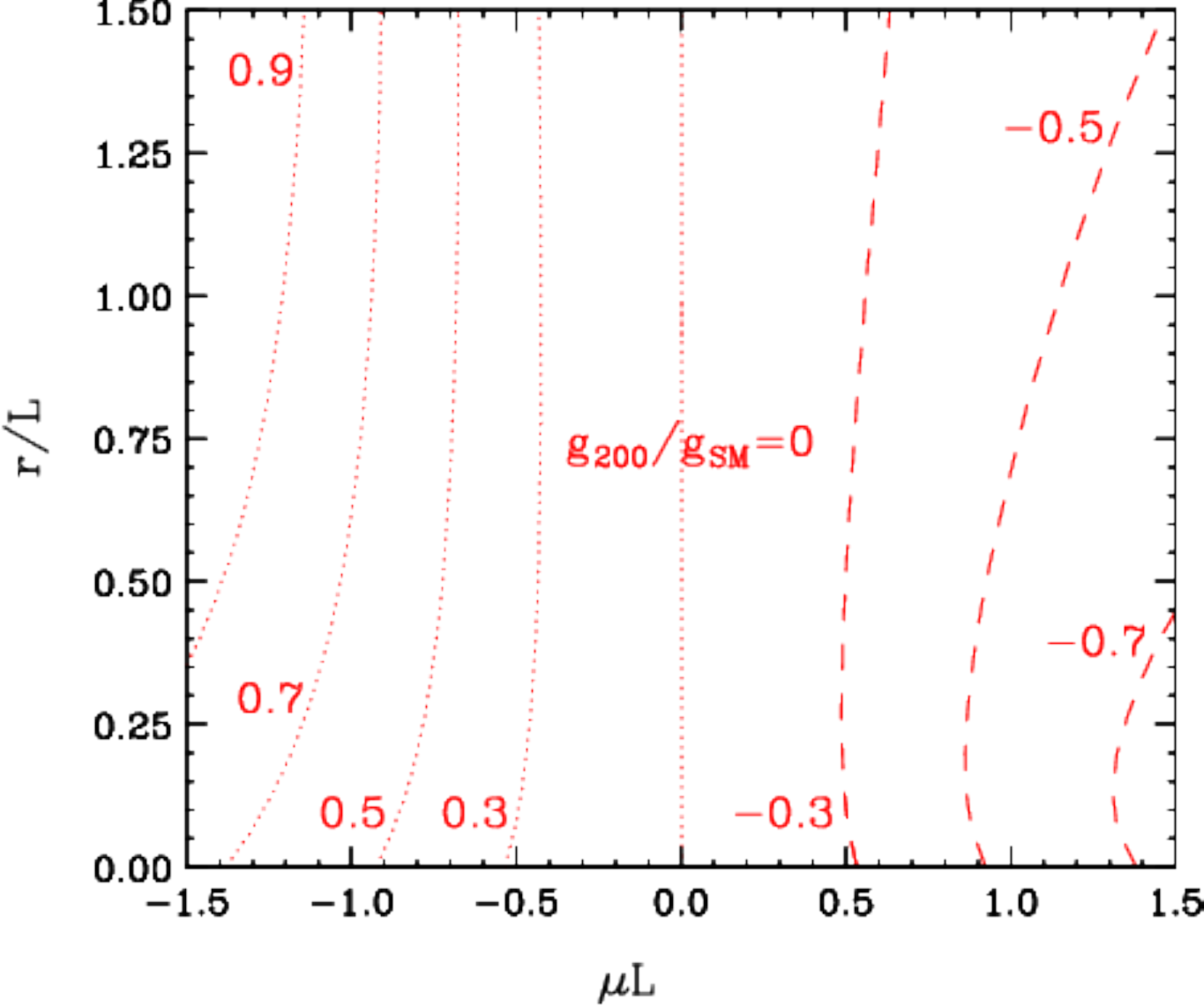, width=0.48\columnwidth}
 } 
\caption{\sl Modified KK couplings: $V_1f_1 f_0$ (top-left), $V_2f_2 f_0$ (top-right), $V_2 f_1 f_1$ (bottom-left), and $V_2 f_0 f_0$ (bottom-right).
\label{fig:couplings} }
\end{figure}

Boundary terms and fermion bulk masses also modify the couplings between KK particles, which are relevant for phenomenological studies. 
Fig. \ref{fig:couplings} (taken from Ref. \cite{Flacke:2013pla}) shows the modified KK couplings: $V_1f_1f_0$, $V_2f_2f_0$, $V_2f_1f_1$ and $V_2f_0f_0$. 
Some comments in couplings are given as follows.
\begin{itemize}
\item KK-parity is always respected so that $V_n f_m f_\ell$ coupling is non-vanishing only when $n+m+\ell$ is an even number so that $(-1)^{n+m+\ell}=1$.
\item KK number is not necessarily respected because of nontrivial boundary conditions and bulk masses.
\item $g_{V_1f_1f_0}/g_{\rm SM}=1$ along $\mu L=0$ but $g_{V_1f_1f_0}/g_{\rm SM}$ becomes larger (smaller) when $\mu L>0 (<0)$, respectively. 
\item $g_{V_2f_2f_0}/g_{\rm SM}=1$ along $\mu L=0$ but there is another branch in the domain with a negative $\mu L$ and positive $r$ which also gives $g_{V_2f_2f_0}/g_{\rm SM}=1$. Above the branch cut, $g_{V_2f_2f_0}/g_{\rm SM}$ becomes larger (smaller) than $1$ for a negative (positive) $\mu L$. Below the branch cut, $g_{V_2f_2f_0}/g_{\rm SM}$ becomes smaller for smaller $r$.
\item The KK number breaking interaction $g_{V_2f_0f_0}$ vanishes for  $\mu L=0$ but is allowed when $\mu L\neq 0$.
\end{itemize}

\section{Precision constraints}

KK-parity odd particles do not contribute to precision measurements at tree-level due to KK-parity but their contribution may appear as loop-corrections, 
such that FCNCs, electroweak precision tests, and constraints on four-fermion contact interactions allow to constrain the scale of new particles.
We discuss implication of UED in each precision measurement. \\[.5cm]
{\it FCNCs:}\\
MUED is minimally flavor violating \cite{Buras:2002ej,Buras:2003mk}. The current bound on the model has been determined in Ref.\cite{Haisch:2007vb} to be $R^{-1}\gsim 600 \mbox{ GeV}$. Generic non-minimal extensions by boundary terms or quark bulk masses can induce FCNCs at the tree-level, 
which are highly constrained due to the absence of an RS-GIM mechanism \cite{Csaki:2010az}. 
Hence boundary terms and bulk masses in the fermion sector are typically assumed to be flavor-blind ({\it cf. e.g.}  Ref.\cite{Flacke:2013pla}),\footnote{Note that boundary terms or bulk masses for right-handed tops are only weakly constrained by flavor physics, though \cite{Datta:2013yaa}.} 
which implies that bounds on the scale of the lightest quark KK resonances are of the order of the MUED bound.\\[.5cm]
{\it Electroweak precision tests:}\\
Loop corrections to the minimal UED model are oblique to a good approximation \cite{Appelquist:2000nn,Appelquist:2002wb} and can be parameterized in terms the Peskin-Takeuchi parameters. The current bound on the MUED model lies at $R^{-1}> 680 \mbox{ GeV}$ \cite{Gogoladze:2006br,Kakuda:2013kba} and results mostly from the loop contributions of the KK top and Higgs modes to the gauge boson propagators. In non-minimal extensions, boundary terms and quark bulk masses lift the degeneracy of the KK mass spectrum. Ref.\cite{Flacke:2013pla} considered a simplified NMUED setup with universal boundary kinetic terms and uniform fermion bulk masses. As was shown there, the electroweak (EW) bounds on the LKP mass can be marginally lowered (by $\mathcal{O}(100 \mbox{ GeV})$) due to the presence of the boundary term through which the higher KK modes contribute less.\footnote{Increasing the mass splitting between the KK top and the KK Higgs allows to reduce the EW bounds further, but this induces signals in 2nd KK mode resonance searches ({\it cf.} Sec.\ref{sec:collider}) which exclude a large mass splitting between KK fermions and gauge bosons if uniform bulk masses and boundary terms are assumed .}\\[.5cm] 
{\it Contact interactions induced by new physics:}\\
NMUED models contain tree-level interactions between SM fermions and even KK mode gauge bosons. At low energies, these interactions induce flavor conserving 4-fermion contact interactions. The constraints from contact interactions yield bounds on the mass of the second KK mode gauge bosons as well as on the KK number violating coupling strength ({\it cf. eg.} Refs.~\cite{Kong:2010xk,Huang:2012kz,Flacke:2013pla}). However, the same type of couplings yield signals for 2nd KK mode resonance searches at LHC which typically imply stronger bounds  ({\it cf.} Sec.\ref{sec:collider}).

\section{Collider phenomenology}\label{sec:collider}

In this section, we provide a brief summary of various collider studies.\\[.5cm]
{\it Vectorlike fourth generation with a discrete symmetry:}\\
Fermion bulk masses allow for a possibility that level-1 fermions are much lighter than all the other KK states. Their masses can be essentially decoupled from the compactification scale, which leads to a new model for vectorlike fourth generations\cite{Kong:2010qd}. This possibility becomes even more important after the LHC ruled out conventional fourth generation models based on chiral fermions. The level-1 right-handed neutrino is a good candidate of LKP DM candidate in this case. The next-to-lightest KK particle (NLKP) fermion would leave distinguishable features in collider experiments such as displaced vertex $\sim 300 {\rm \mu m}$ if the life time is $\sim 10^{-12}{\rm s}$. For instance, the level-1 electron decays through 
$e_1 \to e Z_1^* \to e \bar{\nu}\nu_1$ where $Z_1^*$ is highly off-shell state thus provides a large suppression in the decay width. The doublet KK top decays via $t_1 \to b W_1^*\to b\bar{\ell}\nu_1$ are also possible, if kinematically allowed, while the pair production of the singlet KK top would give $t\bar{t}+\met$ through $Z_1^*\to \nu\nu_1$.\\[.5cm]
{\it Higgs phenomenology:}\\
UED models predict an enhanced Higgs production rate via gluon fusion as well as a modified decay rate for $h\rightarrow \gamma\gamma$ because the top KK modes and (for the decay) $W$ KK modes contribute to these one-loop induced processes \cite{Petriello:2002uu,Belanger:2012mc,Kakuda:2013kba}.  Within the minimal UED model, the current ATLAS and CMS Higgs measurements \cite{ATLASH,CMSH} imply a lower bound of $R^{-1}\gtrsim 460 \mbox{GeV}$ (from ATLAS data) and $R^{-1}\gtrsim 1300 \mbox{GeV}$ (from CMS data) \cite{Datta:2013xwa}.\footnote{The bound dominantly arises from the stronger (weaker) deficit in the $pp\rightarrow h\rightarrow WW^*$ channel observed by CMS (ATLAS).}  In 6D UED models, these bounds are enhanced \cite{Nishiwaki:2011vi,Kakuda:2013kba} due to the larger number of top KK partners. In non-minimal UED models, the constraint can be weakened, however, as the mass of the top KK modes can be raised relative to the LKP, such that the effect of top KK mode loops becomes suppressed \cite{Flacke:2013nta}.\\[.5cm]
{\it Searches for signals with $\met$:}\\
Classical UED search channels are signals with a large missing $p_T/E_T$ which arise when first KK level particles are pair-produced and subsequently cascade-decay into SM particles and the LKP \cite{Cheng:2002ab,Cacciapaglia:2013wha}. These signatures qualitatively resemble R-parity conserving SUSY signals \cite{Cheng:2002ab}. However, in the case of MUED, the KK mass spectrum is compressed. Therefore, the final state SM particles are soft 
and the two LKPs in the final state tend to be back-to-back, implying a small overall $\met$, which makes the signature harder to distinguish from background  compared to searches for SUSY (with non-compressed spectra). For MUED, the most promising search channels are the three and four-lepton + $\met$ signals  \cite{Belyaev:2012ai,Murayama:2011hj} which have an exclusion potential up to $R^{-1}\gtrsim 1.2$ TeV for the current LHC data at 8 TeV \cite{Belyaev:2012ai}.
In non-minimal UED models, the near-degeneracy of the first KK mode masses is lifted, which generically improves the bounds, but the precise bounds (and best-suited search channels) strongly depend on a choice of boundary terms and bulk masses assumed \cite{Chen:2009gz,Datta:2012tv,Datta:2013yaa}. \\[.5cm]
{\it 2nd KK mode resonance searches:}\\
Within the minimal UED model (as well as 6D UED models without boundary terms), Kaluza-Klein number is preserved at tree-level. KK number violating (but KK-parity preserving) interactions -- in particular the interactions between two Standard model particles and one 2nd KK mode particle -- are only induced at loop-level \cite{Cheng:2002iz}. Nevertheless, even in minimal UED, KK number violating interactions are relevant in constraining the model. 2nd KK mode particles can be singly produced and can decay into two Standard Model particles, such that the respective 2nd KK modes have  $Z'$-, $W'$-, or colored resonance-like signatures \cite{Datta:2005zs}. As shown in Ref.\cite{Edelhauser:2013lia}, the absence of any $Z'$-like signals in the di-lepton channel of CMS and ATLAS searches results in a constraint of $R^{-1}>715$~GeV for the MUED model.\\
In non-minimal UED models, non-uniform boundary kinetic terms and/or bulk fermion masses induce KK number violating interactions already at tree-level. Therefore (depending on the size of the bulk and boundary parameters) 2nd KK mode particles can be produced with couplings of the order of the corresponding Standard Model coupling strength ($g_3$ for colored 2nd KK modes, $g_2$ for weakly coupled 2nd KK modes, etc.). Searches of 2nd KK mode gluons in the dijet channel \cite{Park:2009cs,Kong:2013xta} 2nd KK mode $Z$ bosons and photons in the di-lepton channel \cite{Kong:2010xk,Flacke:2013pla}  and 2nd KK mode $W$ bosons in the $l\nu_l$ channel \cite{Kim:2011tq} and the $tb$ channel \cite{Flacke:2012ke} have been performed in various non-minimal UED setups\footnote{Refs. \cite{Park:2009cs,Chen:2009gz,Kong:2010xk,Kong:2010qd,Kim:2011tq,Huang:2012kz,Rizzo:2012rb,Flacke:2012ke,Majee:2013se} study UED models with fermion bulk mass,  \cite{Flacke:2012ke,Datta:2012tv,Datta:2013yaa}  study boundary localized kinetic terms, while Ref. \cite{Flacke:2013pla} considers a non-minimal UED model with uniform boundary kinetic terms and a uniform bulk mass.}. 
Typical NMUED bounds on 2nd KK modes are stronger than the MUED bounds on 2nd KK modes (which are weaker than the projected $\met$ signal bounds), but the currently studied NMUED bounds make various simplifying assumptions (for example universality of the boundary terms).

\section{Dark matter}

UED models provide a variety of dark matter candidates with different features. \\[.5cm]
{\it Dark matter candidates:}\\
Just as in MUED, the lightest KK particle (LKP) is automatically stable due to the KK-parity, thus, if the LKP is neutral, it may serve as a weakly interacting massive particle (WIMP) dark matter with its naturally weak couplings. The KK gauge boson of the hypercharge, often called KK-photon, $\gamma_1$ ( or $B_1$), turns out to be the LKP in MUED and is mostly studied but in the extensions of UED, there are other possible candidates
\cite{Flacke:2008ne,Servant:2002aq,Cheng:2002ej,Kong:2005hn,Burnell:2005hm}. 
Importantly, the non-zero bulk mass of a fermion makes the mass of KK fermions heavier (here we just consider the `heavy' solution) but the brane localized kinetic terms make the mass scale lighter so that the realization of the LKP depends on several parameters. Among others, the notable WIMP candidates are $\gamma_1$ (the KK photon), $W_3^1$ (the KK-Z boson), $H^1$ (the KK Higgs boson),  KK graviton, $\nu_L^1$ (the KK state of the left-handed neutrino) and $\nu_R^1$ (the KK state of the right-handed neutrino)\cite{Kong:2010qd}, which may be realized with different sets of bulk mass parameters and boundary terms. Note that these KK DM candidates qualitatively resemble neutralinos (bino, wino and higgsino), gravitino and sneutrinos in supersymmetric models with different spin states.\\[.5cm] 
{\it Relic density:}\\
The relic abundance of a WIMP particle ($\chi$)  is approximately given by the annihilation cross section ($\sigma_A v$) of the DM candidate 
\begin{eqnarray}
\Omega_{\chi} h^2 \simeq \frac{0.1 {\rm pb} \cdot c}{\langle \sigma_A v\rangle} \, .
\end{eqnarray}
Thus to fit the observed amount $\Omega_{DM} h^2 \simeq 0.11$, one requires $\langle \sigma_A v\rangle \sim 1{\rm pb}$. In NMUED, the degeneracy in mass spectrum is less significant than in MUED, the coannihilation among KK photon and KK fermions may be negligible in the bulk of parameter space.
In case of MUED, the WMAP consistent $R^{-1}$ is about $900$ GeV but with the non-zero bulk mass with $\mu L \sim -0.2, r/L\sim 0$, $R^{-1}=700$ GeV is still allowed \cite{Flacke:2013pla}. With a larger value $r/L\sim 1.2$ and $\mu L\sim -0.3$, the scale is higher $R^{-1}=1300$ GeV. \cite{Flacke:2013pla} 
If non-universal bulk masses are allowed, the branching fraction of $B_1 B_1$ annihilation into different flavors of fermions can vary.  If a larger masses for KK quarks are induced by the bulk mass, e.g., a larger branching fraction to leptons are realized and that may explain the `positron excess' recently found in PAMELA, FERMI and AMS02 \cite{Park:2009cs,Chen:2009gz,Chen:2009zs}.

The other case is that LKP is produced from the decay of NLKP such as $\nu_R^1$ (LKP) from $\nu_L^1 \to \nu_R^1 + X$ where $X=h,Z$. The relic abundance of LKP is determined by the mass ratio between NLKP and LKP and the abundance of NLKP \cite{Kong:2010qd}
\begin{eqnarray}
\Omega_{LKP} = \frac{m_{LKP}}{m_{NLKP}}\Omega_{NLKP}.
\end{eqnarray}
\\[.5cm]
{\it Indirect detection:}\\  The branching fraction of the annihilation process of KK-photon DM to a pair of fermions, $B_1 B_1 \to f \bar{f}$,  is sensitive to the mass of the KK-fermion in the t-channel because the cross section is given as\cite{Park:2009cs,Chen:2009gz,Bertone:2002ms}:
\begin{eqnarray}
\langle \sigma v \rangle_{B_1 B_1 \to f \bar{f}} \propto \frac{m_{B_1}^2}{(m_{B_1}^2 + m_{f_1}^2)^2}.
\end{eqnarray}
If the KK-mass of the fermion is large, the branching fraction becomes small. The KK-mass of a fermion becomes large when a large bulk mass is introduced but the zero mode obtains its mass through usual Higgs mechanism. This phenomenon is important when one consider the indirect signals of LKP DM in space. Especially in the galactic center or local clumpy area or in the sun, DM particles could be accumulated and then annihilation process takes place at the rate, $\Gamma \propto \rho_{DM}^2$. Some of the produced particles, e.g., $e^+, e^-, \nu, p, \bar{p}$ and $\gamma$, propagate toward the earth and can be detected as high energy cosmic ray particles. Unstable particles also contribute to the cosmic ray by cascade decay processes into stable particles. When charged particles are produced by DM annihilation, they radiate off diffuse gamma rays which contribute to the gamma ray flux from DM abundant regions. 

Recent observation of `excess' in cosmic positrons/electrons and non-observation of anti-proton by PAMELA~\cite{Adriani:2008zr}, Fermi-LAT\cite{FermiLAT:2011ab} and AMS02\cite{Aguilar:2013qda} could be interpreted as indirect signatures of KK DM~\cite{Park:2009cs,Chen:2009gz}. The leptophilic property can be realized by the bulk masses of quarks, which provide a sizable suppression factor $\sim 1/m_{f_1}^4$.  For a recent discussion on flavor universal bulk mass on the positron excesses, see Ref. \cite{Gao:2014wga}. \\[.5cm] 
{\it Direct detection:}\\ A boundary term for a boson makes the level-1 KK state lighter. Depending on the parameter values, the LKP can be the level-1 photon ($B_1$), the level-1 $Z$-boson ($~W_3^{(1)}$) but a 1 TeV mass $W_3^{(1)}$ has too small DM abundance thus is not favored \cite{Datta:2013nua}.  The expected signal rate of KK DM with, e.g., Xenon nuclei is still well below the existing limits set by XENON100 for spin-independent scattering \cite{Datta:2013nua}.
Higgs DM in UED models is discussed in Ref. \cite{Melbeus:2012wi}. Direct detection in UED with a general mass spectrum is discussed in Refs. \cite{Servant:2002hb,Arrenberg:2008wy,Arrenberg:2013paa}.

\section{Conclusion}

Models with Universal extra dimensions provide an attractive alternative physics beyond the standard model with rich phenomenology at the LHC and various potential dark matter candidates as shown in Table \ref{ta1}. As an effective theory, the minimal UED is naturally extended to non-minimal UED models allowing the bulk masses for fermions and also brane localized terms, which open additional parameter space of distinctive phenomenological features. In this review, we summarize what we have learned in the direction of NMUED with specific parameter choices under some reasonable but certainly-not-most-general assumptions. Particular set of observables in electroweak precision measurements, collider experiments and dark matter experiments are studied and 
further studies are definitely required to probe the full parameter space. 
Finally we emphasize that non-minimal UED allows a large territory for phenomenological studies.

\begin{table}[t]
\tbl{\label{ta1}Overview on 
the minimal UED model and various extensions. }
{\begin{tabular}{@{}lcccc@{}} \toprule
parameter/model  & MUED & SUED & NMUED & others\\
& $S^1/{\mathbb{Z}_2}$ & $S^1/{\mathbb{Z}_2}$ & $S^1/{\mathbb{Z}_2}$ & $T^2/{\mathbb{Z}_4}$ ...\\
\colrule
Bulk Mass  & 0 & $\mu_{\Psi} \theta(y)$  & $\mu_{\Psi} \theta(y)$ & usually $0$  \\
Boundary terms & 0 & 0 & non-zero & -- \\
LKP & $B_1$ & $B_1, n_1..$ & $B_1, Z_1..$ & $s=0$ photon ..\\
\botrule 
\end{tabular}}
\end{table}

\section*{Acknowledgments}

TF is supported by the National Research Foundation of Korea (NRF) grant funded by the Korea government (MEST) N01120547. He would also like to thank CERN for hospitality, where part of this work as been done.
KK is supported in part by the US DOE Grant DE-FG02-12ER41809 and by the University of Kansas General Research Fund allocation 2301566. SC is supported by Basic Science Research Program through the National Research Foundation of Korea funded by the Ministry of Science, ICT $\&$ Future planning (NRF-2011-0029758) and (NRF-2013R1A1A2064120).

\end{document}